\documentclass{aa}  
\usepackage{graphicx}
\usepackage{txfonts}
\usepackage{natbib}
\usepackage[german,english]{babel}
\begin{document}
   \title{GMASS Ultradeep Spectroscopy of Galaxies at z$\sim$2. I. The
stellar metallicity \thanks{ %
   Based on observations obtained at the ESO Very Large Telescope
   (VLT) as part of the Large Programme 173.A--0687 (the Galaxy Mass
   Assembly ultradeep Spectroscopic Survey).}}

   \subtitle{}

   \author{C.~Halliday
          \inst{1}
          \and
          E.~Daddi\inst{2}\
          \and
          A.~Cimatti\inst{3}\
          \and
          J.~Kurk\inst{4}\
          \and
          A.~Renzini\inst{5}\
          \and
          M.~Mignoli\inst{3}\
          \and
          M.~Bolzonella\inst{3}\
          \and
          L.~Pozzetti\inst{3}\
          \and
          M.~Dickinson\inst{6}\
          \and
          G.~Zamorani\inst{3}
          \and
          S.~Berta\inst{7}\
          \and
          A.~Franceschini\inst{7}\
          \and
          P.~Cassata\inst{8}\
          \and
          G.~Rodighiero\inst{7}\
          \and
          P.~Rosati\inst{9}\
          }

   \offprints{C. Halliday}

   \institute{Osservatorio Astrofisico di Arcetri, Largo Enrico Fermi 5, I-50125 Firenze, Italy\\
              \email{halliday@arcetri.astro.it}
         \and
             Laboratoire AIM, CEA/DSM - CNRS - Universit\`e Paris Diderot, DAPNIA/SAp, Orme des Merisiers, 91191 Gif-sur-Yvette, France
         \and
             Dipartimento di Astronomia, Alma Mater Studiorum, Universit{\'a} di Bologna, Via Ranzani 1, I-40127 Bologna, Italy
         \and
             Max-Planck-Institut f{\"u}r Astronomie, K{\"o}nigstuhl 17, D-69117 Heidelberg, Germany
          \and
             Osservatorio Astronomico di Padova, vicolo dell'Osservatorio 5, I-35122 Padova, Italy
         \and
             National Optical Astronomy Observatory, 950 North Cherry Avenue, Tuscon, AZ 85719, USA
         \and
             Dipartimento di Astronomia, Universit{\'a} di Padova, vicolo dell'Osservatorio 3, I-35122 Padova, Italy
          \and
             Laboratoire d'Astrophysique de Marseille (UMR 6110), CNRS, Universit\`e de Provence, BP 8, 13376 Marseille Cedex 12, France
          \and
             European Southern Observatory, Karl-Schwarschild-Str.~2, D-85748 Garching, Germany
             }

   \date{Received ---; accepted ---}

 
  \abstract
   {Galaxy metallicities have been measured to redshift z$\sim$2 by
   gas-phase oxygen abundances of the interstellar medium using the
   R$_{23}$ and N2 methods. Galaxy stellar metallicities provide
   crucial data for chemical evolution models but have not been
   assessed reliably much outside the local Universe.}
   {We determine the iron-abundance, stellar metallicity of
   star-forming galaxies at redshift z$\sim$2, homogeneously-selected
   and observed as part of the Galaxy Mass Assembly ultra-deep
   Spectroscopic Survey (GMASS).}
   {We compute the equivalent width (EW) of a rest-frame
   mid-ultraviolet (mid-UV), photospheric absorption-line index, the
   1978~\AA\ index, found to vary monotonically with stellar
   metallicity by Rix, Pettini and collaborators (R04), in model
   star-forming galaxy (SFG) spectra created using the
   theoretical massive star models of Pauldrach and coworkers, and
   the evolutionary population synthesis code Starburst99. The
   1978~\AA\ index is sensitive to Fe III transitions and measures the
   iron-abundance, stellar metallicity. To accurately determine the
   1978~\AA\ index EW, we normalise and combine 75 SFG spectra from
   the GMASS survey to produce a spectrum corresponding to a total
   integration time 1652.5 hours (and a signal-to-noise ratio
   $\sim$100 for our~1.5~\AA~binning) of FORS2 spectroscopic
   observations at the Very Large Telescope.}
   {We measure a iron-abundance, stellar metallicity of
   log$(Z/Z_\odot) = -0.574\pm0.159$ for our spectrum
   representative of a galaxy of stellar mass
   9.4$\times$10$^{9}$~M$_\odot$ assuming a Chabrier initial mass
   function (IMF). We find that the R04 model SFG spectrum for
   log$(Z/Z_\odot) = -0.699$ solar metallicity provides the best
   description of our GMASS coadded spectrum. For similar galaxy
   stellar mass, our stellar metallicity is $\sim0.25$ dex lower than
   the oxygen-abundance, gas-phase metallicity quantified by
   Erb and collaborators (E06) for UV-selected star-forming
   galaxies at $z=2$.}
   {We measure the iron-abundance, stellar metallicity of
   star-forming galaxies at redshift z$\sim$2 by analysing the
   1978~\AA\ index in a spectrum created by combining 75 galaxy
   spectra from the GMASS survey. We find that our measurement is
   $\sim$0.25 dex lower than the oxygen-abundance gas-phase
   metallicity at similar values of galaxy stellar mass. We conclude
   that we are witnessing the establishment of a light-element
   overabundance in galaxies as they are being formed at redshift
   z$\sim$2. Our measurements are indeed reminiscent of the
   $\alpha$-element enhancement seen in the likely progenitors of
   these starburst galaxies at low-redshift, i.e. galactic bulges and
   early-type galaxies.}


   \maketitle
%

\section{Introduction}

Galaxies are believed to form by the merging of dark matter haloes,
the isolated collapse of a gas cloud and subsequent gas accretion, or
a combination of these scenarios. Photometric and spectroscopic
surveys measure the rate of galaxy mass assembly and the variation in
galaxy star formation rates with redshift
\citep{lil96,mad96,mau04,cim04}.

Galaxies are unlikely to evolve as closed boxes but exchange gas,
metals and stars with their environment. Following the early work of
\cite{lar74} cosmological models propose that mass outflows and winds
driven by supernovae and AGN activity suppress star formation (via
so-called ``feedback'') increasingly effectively in galaxies of
increasingly lower mass, and enrich the interstellar medium (ISM) with
products of stellar nucleosynthesis
\protect\citep{spr03,ben03,del04,cro06,bow06,brt07,fin07}. Such winds
are required for example to prevent the overproduction of galactic
low-mass stars and describe the observed metal-enrichment history of
the Universe. In the local Universe and to a redshift of z$\sim$2 a
``mass-metallicity relation'' is observed such that more massive
star-forming galaxies are more
metal-rich~\citep{leq79,tre04,sav05,erb06,lee06}. \cite{tre04}
(hereinafter T04) found a tight correlation between the gas-phase
metallicity and stellar mass of 53,400 star-forming galaxies of
redshift 0.005$ < $z$ < $0.25, using imaging and spectroscopic data
from the Sloan Digital Sky Survey (SDSS). Their relation has a
1$\sigma$ scatter from the median of only 0.1 dex and is linear for
stellar masses 8.5 $<$ log(M/M$_\odot$) $<$ 10.5. At higher
redshift~\citet{fin07} were able to more accurately reproduce the gas
phase mass-metallicity relation at z$\sim$2~using ``momentum-driven
winds'' \citep{mur05,opp06} instead of constant velocity winds
traditionally proposed. ``Momentum-driven winds'' remove metals into
the intergalactic medium by radiation pressure as photons are absorbed
and scattered by dust, and supernovae and galactic winds. Unlike
thermal energy, momentum cannot be radiated away and ``momentum-driven
winds'' can transport metals to significant distances outside the
galaxy.

Measuring galaxy metallicity as a function of redshift probes the rate
of chemical enrichment. In the local Universe metallicity is measured
using both spectroscopic absorption and emission lines. \cite{gal05}
found good agreement between stellar and gas-phase metallicities using
SDSS galaxy spectra for which both absorption and emission lines were
measurable. Faber and collaborators at Lick Observatory pioneered
measurement of luminosity-weighted stellar age and metallicity of
nearby early-type galaxies by developing the ``Lick system'' of
spectroscopic absorption-line indices
\citep{bur84,fab85,gon93,gor93,guy94,wor97,tra98} for the rest-frame
optical wavelength range $\sim$4000$-$6200\AA, and the single-burst
evolutionary population synthesis models of \cite{cla94}. The ``Lick
system'' was further developed by \cite{tra00} and Thomas, Maraston
and collaborators \citep{tho03,tho04} to account for the effects of
light-element overabundance ratios on Lick/IDS index measurements
using respectively the empirical relations of \cite{tri95}, and the
theoretical models of \cite{mar05} and calibrations of \cite{kor05}.

At high-redshift emission lines are more readily detected than
absorption lines and galaxy metallicity has been measured using
primarily the oxygen-abundance, gas-phase metallicity of the ISM of
star-forming galaxies, using the R$_{23}$ \citep{pag79} and N2
\citep{sto94,rai00} methods
\citep{kza99,kko00,cli01,lil03,kob04,sha04,mai05,erb06,lam06,mai06}.
At z$>$1 the gas metallicity of the ISM has been constrained using
the~CIV$\lambda$1550~\citep{meh02} and MgII and FeII absorption lines
\citep{sav04}. Pettini and collaborators have measured the chemical
abundances of the ISM using rest-frame UV absorption lines in a
spectrum of the gravitationally-lensed Lyman-$\alpha$ galaxy MS
1512-cB58 at redshift z$\sim$2.73 \citep{pet02} and damped
Lyman-$\alpha$ systems \citep{pet02a}. Stellar metallicity data are
required to fully test chemical evolution models. \cite{dem04}
provided a first constraint of stellar metallicity for a small sample
of very massive ($>10^{11}M_\odot$) $z\sim2$ galaxies from the K20
survey \citep{cim02,ema04}. They showed tentative evidence for solar
or supersolar metallicity but with large uncertainties due to
limitations in the achieved S/N ratio. \protect\cite{rix04} compared
the iron-abundance, stellar metallicity for MS 1512-cB58 at z=2.73,
and Q1307-BM1163 at z=1.411, with previously determined light-element
abundances (magnesium, silicon and sulphur) and iron-abundance
gas-phase metallicities of the ISM. For MS 1512-cB58, the
iron-abundance, stellar metallicity is a factor of two higher than the
light-element abundance gas-phase metallicities. The iron-abundance
gas-phase metallicity is a factor of four lower than the light-element
abundance metallicities: this is attributed to dust depletion of the
ISM and the time delay in release of products of supernovae Ia
\protect\citep{pet02}. For Q1307-BM1163, the iron-abundance, stellar
metallicity is a factor of two higher than the oxygen-abundance,
gas-phase metallicity measured using the N2 method. \cite{erb06}
compared two rest-frame mid-UV spectra (obtained for four of the six
bins of stellar mass examined in their mass-metallicity relation
analysis i.e. two lowest and two highest of six bins of mass) with
star-forming galaxy spectra generated using the Starburst99
\citep{lei99} model. \cite{erb06} found good agreement between the
gas-phase metallicity oxygen abundance determined using the N2 method
and the metallicity of the model spectrum that best described the
rest-frame mid-UV spectra as judged by eye.

We present the first robust measurement of iron-abundance, stellar
metallicity for mid-IR selected star-forming galaxies (SFGs) at
z$\sim$2. We measure a photospheric absorption-line index, defined by
\citet{rix04} (hereinafter R04) in the rest-frame mid-ultraviolet
galaxy spectrum (mid-UV) that is dominated by the light of massive
stars. R04 modelled star-forming galaxy spectra using the evolutionary
population synthesis models Starburst99 (\citet{lei99}; hereinafter
L99) and non-local thermal equilibrium (NLTE) stellar atmosphere
models of OB stars \citep{pau01}. Using stellar atmosphere models, the
wavelength coverage of theoretical star-forming galaxy (SFG) models
was extended redward to $\sim$2100~\AA. Star-forming galaxy model
spectra were generated for five values of metallicity,
0.05,~0.2,~0.4,~1.0,~and~2.0 solar, assuming a continuous star
formation history and a Salpeter IMF. R04 found that the equivalent
width (EW) of an absorption-line system, centred on 1978~\AA~within an
Fe~III transition blend, varied monotonically with metallicity, after
100 Myr of star formation. They defined the EW of the
``1978~\AA~index'', EW(1978), to have limits of 1935~\AA~and
2020~\AA~avoiding the~Al~III $\lambda$$\lambda$1855, 1863 interstellar
lines, the nebular emission line C~III$]\lambda$1909 and other weaker
lines.

We measure the 1978~\AA~index in a spectrum created by adding 75
star-forming galaxy spectra from the Galaxy Mass Assembly ultradeep
Spectroscopic Survey (GMASS) data corresponding to over 1652.5 hours
of integration time with FORS2 at the VLT. Given the difficulty of
measurement of the 1978~\AA~index and the limited range of galaxy
stellar masses, we do not analyse spectra as a function of stellar
mass. It would indeed be challenging to achieve this measurement for
massive galaxies alone, because the most massive galaxies are fainter
in the rest-frame mid-UV (e.g. \citealt{ema04,kon06}) and they are of
course rarer (cf.~Figure \ref{hist}).

In Section \ref{data} we provide a brief description of the GMASS
survey. In Sections \ref{selection} and \ref{coadd}, we describe
respectively, the selection of, and the combination of GMASS
star-forming galaxy spectra. In Section \ref{metallicity} we outline
the measurement of the 1978~\AA~index and its error, and contrast our
observed data with the R04 star-forming galaxy model spectra. In
Section \ref{discuss} the 1978~\AA~index measurement is compared with
predictions of galaxy stellar and gas metallicity from the
cosmological simulations of \citet{fin07}, and the oxygen-abundance
gas-phase metallicities of \citet{erb06} and we briefly discuss our
results and conclusions.

We use AB photometric magnitudes, and assume a WMAP cosmology with
$\Omega_\Lambda, \Omega_M = 0.73, 0.27$, and $h = $H$_0$[km s$^{-1}$
Mpc$^{-1}$]$/100=0.71$ \citep{spe03}. We assume a Chabrier initial
mass function (IMF) \protect\citep{cha03} from 0.1 to 100
$M_\odot$. Whether the IMF is universal between galaxies and at
different redshift is a subject of ongoing debate. The empirical
measurement of the Galactic IMF is not well determined and the
uncertainty increases for the individual kinematical components (disk,
bulge and thick disk) of the Galaxy \protect\citep{cha03}, and to
high-redshift. For a review of the observational constraints provided
by measurement of the mass-to-light ratios of nearby galaxies, the
redshift evolution of the Fundamental Plane and galaxy cluster,
element-abundances, see \protect\cite{ren05}. In this paper, the
Chabrier IMF is chosen to provide a reference for all data shown and
our conclusions are independent of this choice.


\section{The GMASS Survey}\label{data}

GMASS (``Galaxy Mass Assembly ultra-deep Spectroscopic Survey''
\footnote{http://www.arcetri.astro.it/$\sim$cimatti/gmass/gmass.html})
is an ESO VLT Large Program project based on data acquired using the
FORS2 spectrograph. A complete description of GMASS photometric
observations, source detection, target selection, determination of
photometric redshifts, mask preparation and spectroscopic observations
and data reduction are provided by Kurk et al.~(in preparation).

The project's main science driver is to measure the physical
properties of galaxies at redshifts $1.5<z<3$ a critical range in the
mass assembly of massive galaxies. Spectroscopy allows the measurement
of reliable galaxy redshifts, stellar masses, star formation rates and
metallicities. The uniqueness of GMASS is its 4.5$\mu$m selection that
provides two major benefits: (1) it detects the peak of the stellar
SEDs (for $\lambda_{rest}=1.6\mu$m) redshifted into the 4.5$\mu$m band
for $z>1.4$, and (2) it is sensitive to the rest-frame mid-IR
emission, i.e. to stellar mass, up to z$\approx$3. The stellar mass
completeness limits are $\log$(M/M$_\odot) \approx$ 9.8, 10.1, and
10.5 for $z=1.4$, $z=2$, and $z=3$, respectively (for $m_{4.5} < 23.0$
and a Chabrier IMF). These mass limits allow mass assembly to be
tracked for the precursors of today's massive galaxies to the most
massive galaxies at $z\geq1.4$.

GMASS target selection was completed for the GOODS-South
field\footnote{http://www.stsci.edu/science/goods} from a region of
$6.8\times6.8$ arcmin$^2$ chosen to match the FORS2 spectrograph
field-of-view. All sources with {\sl Spitzer Space Telescope} + IRAC
data (Dickinson et al. in preparation), were selected to a limiting
magnitude of $m_{4.5} < 23.0$ (2.3\,$\mu$Jy). Finally a cut in
photometric redshift of $z_{phot} > 1.4$ and two cuts in the optical
magnitudes ($B < 26.5$, $I < 26.5$) were applied.

Spectroscopic observations were completed for integration times of up
to 32 hours, using either the blue 300V grism (for observed
wavelengths 4000--6000 \AA), or the red 300I grism (6000--10000 \AA)
depending on target photometric properties. The spectroscopic slit
width was 1 arcsecond. Overall the GMASS survey had an excellent
spectroscopic redshift measurement success rate of $\sim$85\%. For
objects without GMASS spectroscopic redshifts we adopted data from the
literature where available, or photometric redshifts derived from our
analysis using the publicly available optical (HST+ACS) \citep{gia04},
near-IR (VLT+ISAAC) (Retzlaff et al. in preparation) and IRAC images
of the GOODS-South field (Dickinson et al. in preparation).

%

\section{Selection of galaxy spectra}\label{selection}

Specifically for the present analysis, all spectroscopic data
collected for the GMASS survey were visually classified as passive
galaxy, star-forming galaxy or stellar spectra, or (if a spectrum had
too low signal-to-noise) to have an uncertain spectral type.

\begin{figure*} 
\centering
\includegraphics[width=17.0cm]{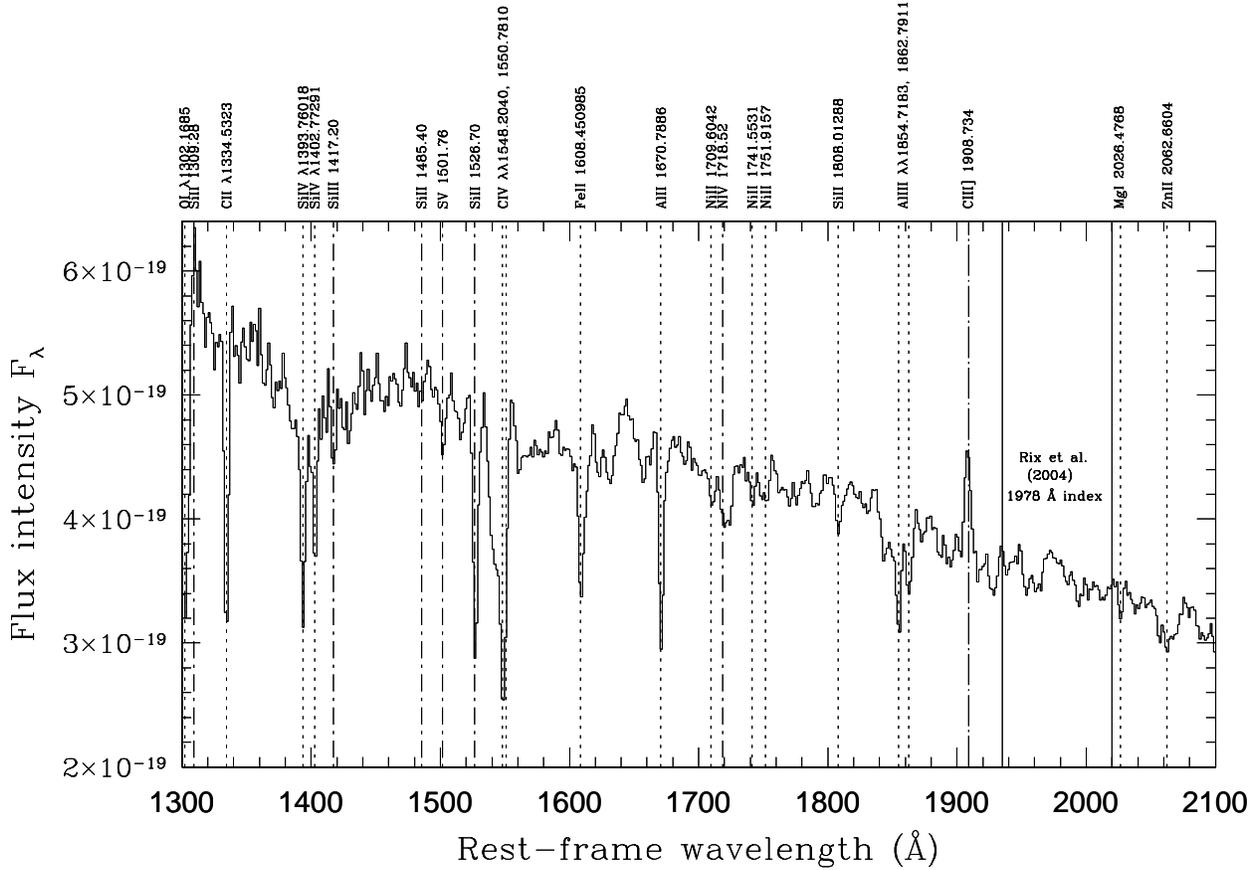}
\caption{\label{coadd_spec} 
We provide an average-combined spectrum of the 75 GMASS star-forming
galaxy spectra selected for the present analysis. Dotted lines
indicate interstellar rest-frame mid-UV absorption lines; dot-short
dashed lines mark photospheric absorption lines; and the dot - long
dashed line labels CIII$] \lambda$ 1908.734 emission. Where two lines
are close in wavelength and correspond to a single element, one label
is shown. The R04 1978~\AA~index equivalent width wavelength range is
delineated by straight solid black lines at 1935~\AA~and 2020~\AA.}
\end{figure*}

\begin{figure*}
\centering
\includegraphics[width=5.5cm]{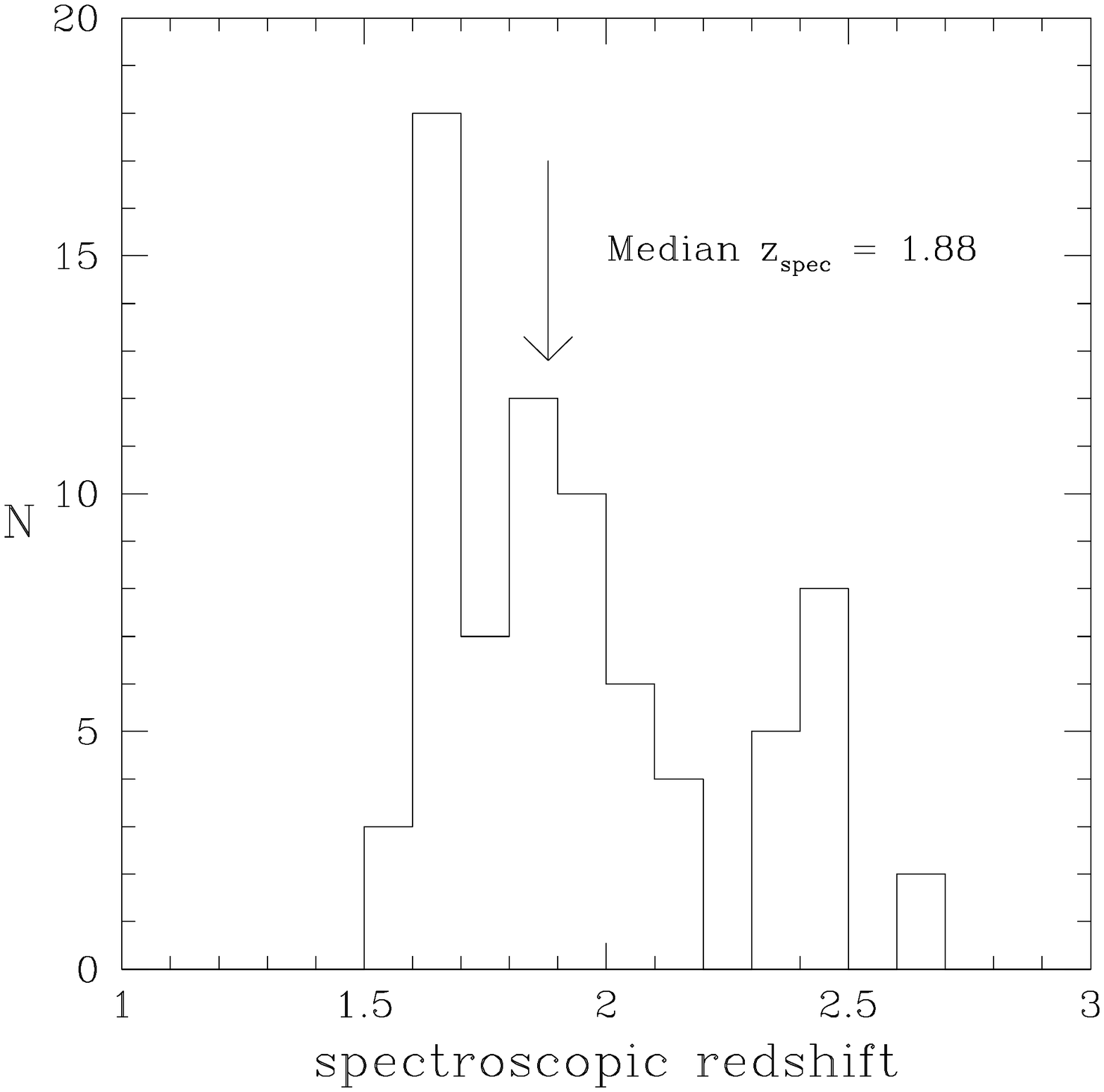}
\includegraphics[width=5.5cm]{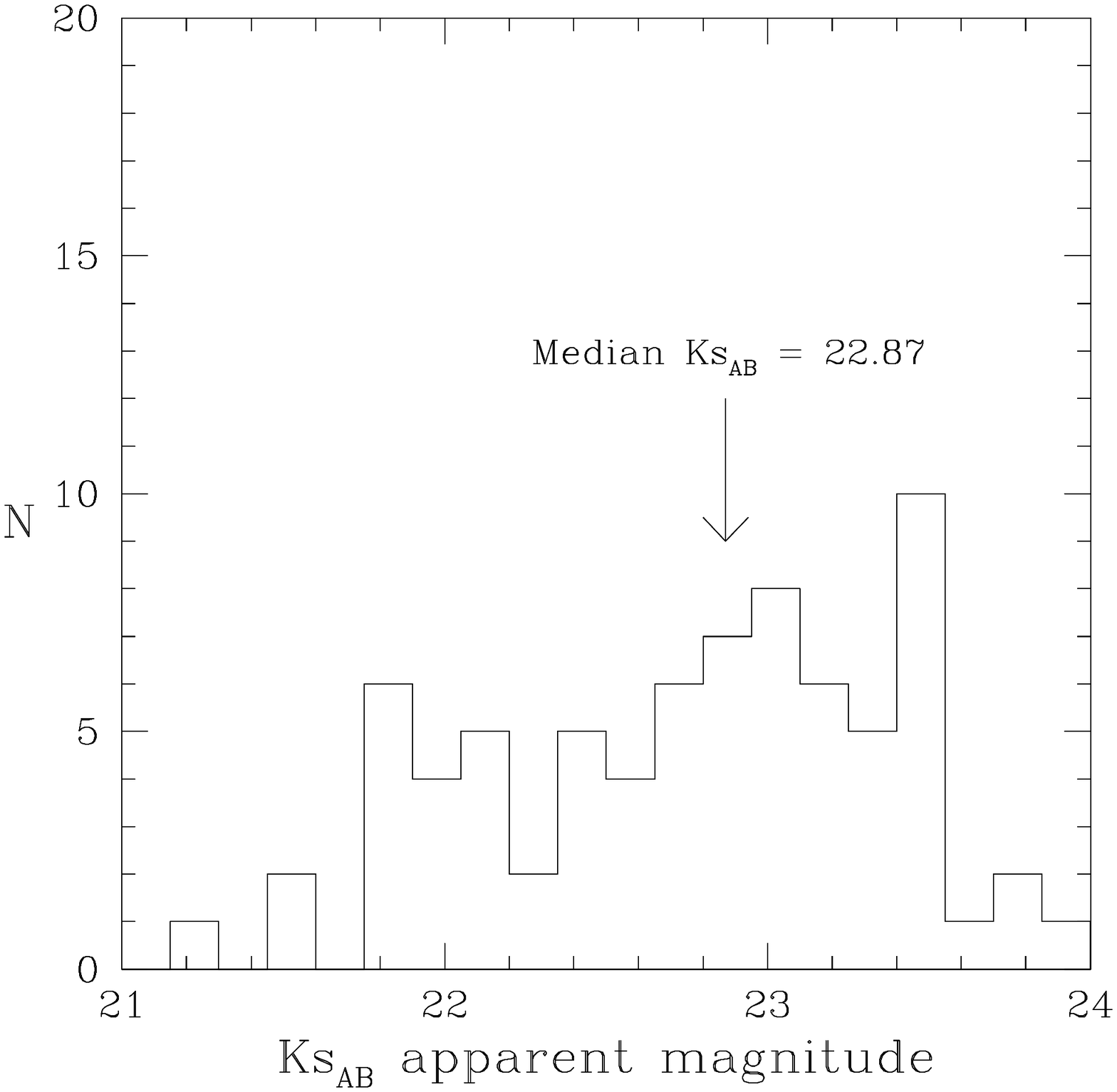}
\includegraphics[width=5.5cm]{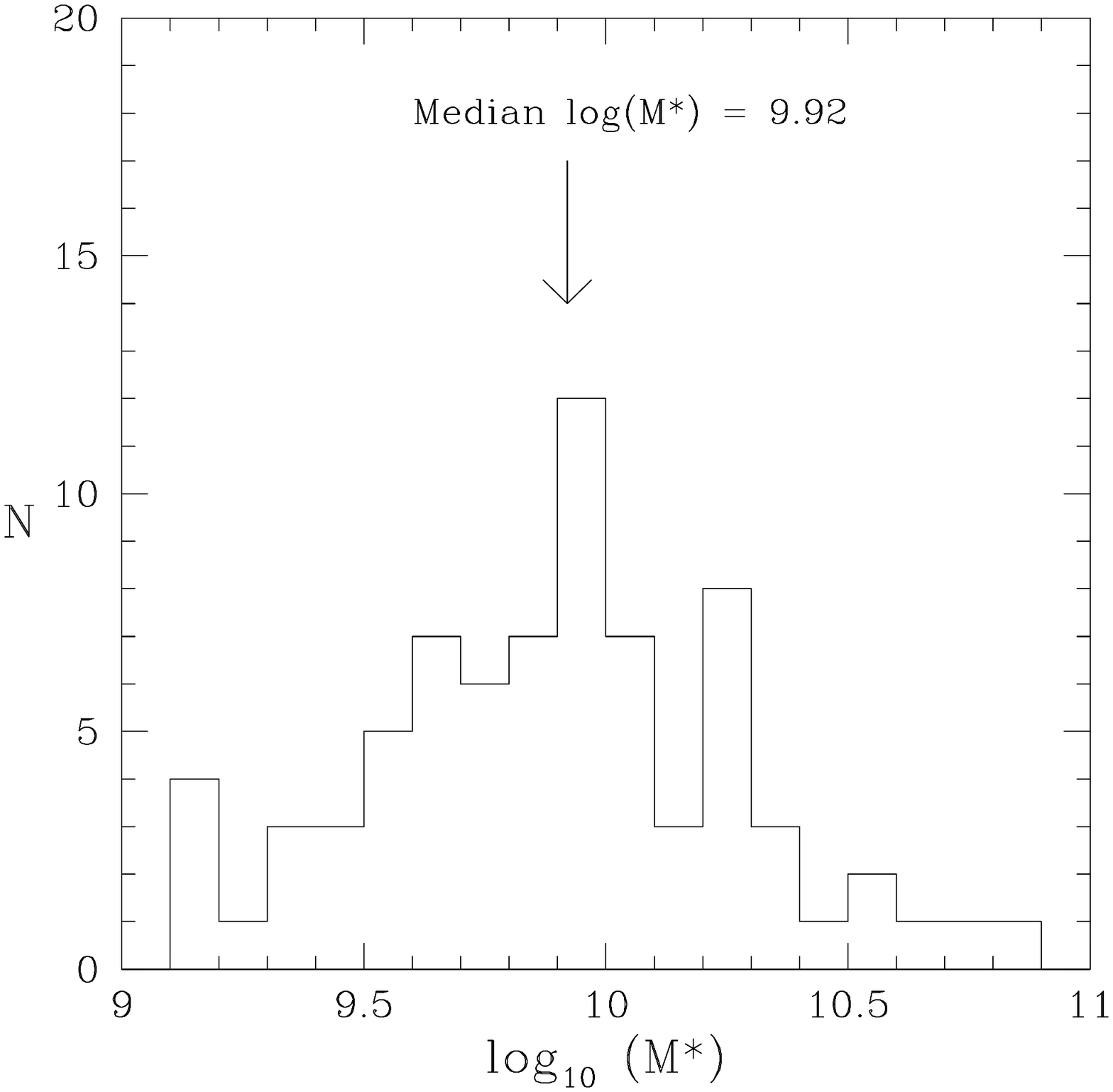}
\caption{\label{hist}
The distribution of GMASS spectroscopic redshifts, GOODS K$_{S}$
magnitudes and GMASS stellar masses of the 75 galaxies for which GMASS
star-forming galaxy spectra are coadded in this analysis.}
\end{figure*}

A spectrum was introduced into our analysis if classified as a
star-forming galaxy (SFG) spectrum and if its rest-frame wavelength
range included 1700-2100~\AA. In Figure \ref{hist} we present the
distribution of redshifts, K$_{S}$ apparent magnitudes and galaxy
stellar masses of the 75 GMASS spectra selected. In Figure
\ref{coadd_spec} we show a typical SFG spectrum created by simply
average-combining all 75 GMASS SFG galaxy spectra chosen.

In Table \ref{med_phys}, we indicate the median physical properties of
the 75 selected galaxies i.e. for photometric data from the GOODS
project, the apparent $B, V, I$ and $z$ photometric magnitudes derived
using HST ACS imaging for the filters F435W (b$_{goods}$), F606W
(v$_{goods}$), F814W (i$_{goods}$), and F850LP (z$_{goods}$)
respectively \citep{gia04}, the $J, H, K_{S}$ photometric magnitude,
and absolute $B, I$ and $K$ magnitude; photometric and spectroscopic
redshift; the best-fit galaxy age in Gyr, star formation rate in solar
masses per year, dust extinction in magnitudes, and stellar mass in
solar masses. Galaxy stellar mass, age, star formation rate, and dust
extinction were measured by fitting the evolutionary population
synthesis models of \cite{bru03} to galaxy spectral energy
distributions SEDs inferred from photometric data gathered for GMASS.
The model assumptions were a Chabrier IMF, solar metallicity, and a
star formation rate (SFR) proportional to $e^{\frac{-t}{\tau}}$ where
$\tau$ is the star formation characteristic timescale, and $t$ the
time elapsed since the onset of the current episode of star formation
(Pozzetti et al.~in preparation).

\begin{table*}

\caption{\label{med_phys} The median galaxy physical properties of 75
GMASS spectra combined to produce a spectrum used to measure the 1978
\protect\AA~index. Data presented are apparent $B, V, I$ and $z$, $J,
H, K_{S}$ photometric magnitudes, absolute $B, I$ and $K$ magnitude
(absB, absI and absK), photometric (z$_{phot}$) and spectroscopic
(z$_{spec}$) redshift, best-fit galaxy age in Gyr, dust extinction
A$_{V}$, star formation rate in solar masses per year, and stellar
mass in solar masses.}
\centering
\begin{tabular}{ccccccccccccccccc}     
\hline
b$_{goods}$ & v$_{goods}$ & i$_{goods}$ & z$_{goods}$ & J & H &  Ks  &  absB  &  absI  & absK & z$_{phot}$ & z$_{spec}$ & age(Gyr) & A$_{V}$ & SFR(M$_\odot$/yr) & Mass(M$_\odot$) \\ 
\hline
 24.63 & 24.41 & 24.02 & 23.82 & 23.34 & 23.16 & 22.87 & -21.44 & -21.96 & -22.29 & 1.91 & 1.88 & 0.18 & 0.80 & 26.58 & 8.32$\times$10$^{9}$ \\
\hline
\end{tabular}
\end{table*}

%

\section{Coaddition of star-forming galaxy spectra}\label{coadd}

To be able to measure the 1978~\AA\ index and the galaxy
iron-abundance, stellar metallicity, we carefully normalised and
combined all 75 SFG spectra selected in Section \ref{selection}.

For all observed wavelengths, the flux-calibrated intensity of each
spectrum was deredshifted to the galaxy rest-frame using the GMASS
spectroscopic redshift, and then divided by the median value
of flux. A spectrum was created for the wavelength range
1700-2100~\AA~by fitting a natural spline function at regular
intervals of 1.5~\AA. A linear function was then fitted to and divided
into this spectrum to take out a remaining slope in the spectral
shape.

Due to the wide range of signal-to-noise ratios of the 75 SFG spectra
(because of the diverse range of galaxy magnitudes and colours) and to
ensure the most precise 1978~\AA~index measurement, we optimised the
signal-to-noise ratio of the coadded spectrum using weights. Each
spectrum was ``weighted'' by a measure of its signal-to-noise ratio
and, as a function of wavelength, by the relative contribution of
background night sky to the spectrum flux. To assess the weight due to
signal-to-noise ratio of each normalised spectrum, we calculated the
semi-interquartile range (s.i.q.r.) of flux in the wavelength range
1920-2050~\AA~and then its quadrature, and subtracted in quadrature a
term (0.03) estimated to represent the contribution of intrinsic
spectrum shape to the calculated s.i.q.r. value. A background sky
spectrum was obtained by average-combining rows containing
predominantly sky spectra in a CCD galaxy spectrum image. At all
wavelengths the background sky flux was divided by its median
value. The weight applied to each individual spectrum during
combination was the product of the inverse of this final background
sky spectrum, and the inverse of the weight due to spectrum
signal-to-noise ratio calculated as described above. A stellar mass of
9.4$\times$10$^{9}$~M$_\odot$ was calculated for the final coadded
spectrum by weighting the GMASS stellar mass for each of the 75
galaxies by the identical weights applied to each spectrum during
spectrum combination.


\section{Stellar metallicity measurement and error}\label{metallicity}

The 1978~\AA\ index has a typical EW of 2-7\AA\ \citep{rix04} and its
measurement is easily affected by a few deviant pixels in particular
inside the pseudo-continua wavelength intervals defined by R04. In
this Section we describe the measurement of the stellar metallicity
and its error, and assess the solidity of our results using
independent methods.

Firstly however we consider the appropriateness of the R04 predictions
for our observed data. The R04 empirical calibration between 1978~\AA\
index EW and metallicity (Equation 8 of R04) is defined for specific
model assumptions and a particular spectral resolution. The R04 models
are applicable to galaxies that have been undergoing bursts of
star-formation for $\geq$ 100 Myr, a reasonable assumption for
star-forming galaxies at redshift 2
\protect\citep{dad04,sha04,erb06}. The rest-frame spectral resolution
of the R04 model predictions is 2.5~\AA; this differs significantly
from our rest-frame spectral resolution of 3.8~\AA\ FWHM.

To be able to reliably measure stellar metallicity we derived a new
empirical calibration between the 1978~\AA\ index and metallicity for
our rest-frame spectral resolution of $\sim$3.8~\AA\ FWHM. To bring
each R04 model spectrum to our resolution, we convolved each spectrum
with a Gaussian function of width 2.86~\AA\ FWHM. We measured the
1978~\AA\ index using an approach identical to the method described
for our observed spectrum below in Section \ref{index}.

\begin{figure} 
\centering
\includegraphics[width=9.0cm]{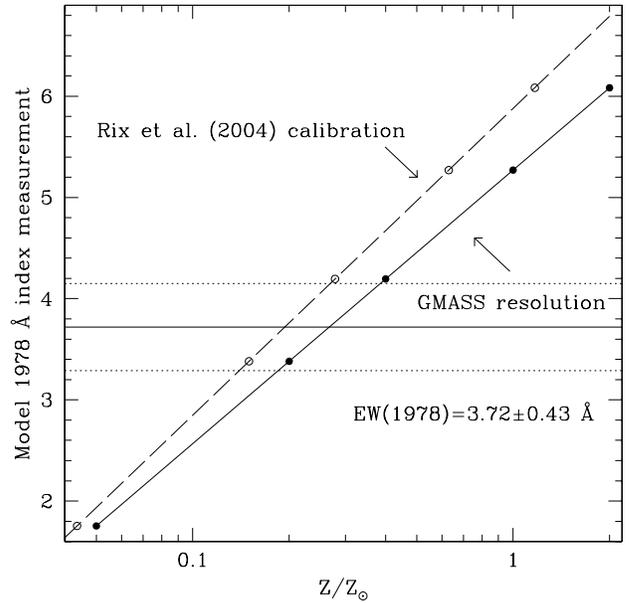}
\caption{\label{rev} Our revised 1978~\AA\ index -
metallicity calibration (solid line) is compared with the original R04
relation (long dashed line). Our 1978~\AA\ index measurement and error
(3.72$\pm$0.43~\AA) are indicated by a horizontal solid and two
horizontal dotted lines respectively. Filled circle symbols indicate
the 1978~\AA\ index measurements for each of the five R04 model
spectra broadened to our rest-frame spectral resolution and their five
values of stellar metallicity. Unfilled circle symbols show the
corresponding metallicity implied by the original R04 relation for the
1978~\AA\ index values measured for each broadened spectrum.}
\end{figure}

Our revision of the R04 calibration is identical in form to Equation 8
of R04 i.e.
\begin{equation}\label{equ}
log (\frac{Z}{Z_{\odot}}) = C \cdot EW(1978) + D
\end{equation}

\noindent but our values of the coefficients C and D are different
i.e. for an EW(1978) $\leq$ 6.0~\AA, we find C = 0.37 and D =
$-$1.95. In Figure \ref{rev} we directly compare the original R04
1978~\AA\ index - metallicity calibration, and our revised
calibration. This figure illustrates that for a given EW(1978)
measurement, our revised calibration implies a higher value of
metallicity than the original R04 calibration. We use our revised
calibration to determine the galaxy iron-abundance, stellar
metallicity.


We remark that to ascertain our 1978~\AA\ index EW and its error and
complete Figures \ref{chisq_model}, \ref{r04model} and
\ref{dave_models}, a small positive offset of 1~\AA\ was applied to
the dispersion axis of the coadded spectrum. This offset was necessary
to optimise the match between our coadded spectrum and the R04 model
spectra i.e.~the wavelengths of the faint absorption-line spectrum
inside 1935--2020\AA. The offset is similar to our wavelength
calibration uncertainty and corresponds to a subfraction of a pixel in
the GMASS spectroscopic data. If the offset had not been applied our
1978~\AA\ index measurement would have changed by far less than the
error we estimate.

\subsection{The 1978~\AA\ index measurement}\label{index}

We measured the 1978~\AA~absorption-line index equivalent width (EW)
as defined by R04 in our coadded spectrum using fixed limits
of~1935~\AA~and 2020~\AA\ and by fitting the continuum using 6 of the
narrow (typically 3-5 \AA\ wide) pseudo-continua intervals for the
wavelength range 1700-2100~\AA.~For our 1978~\AA~index EW measurement
of 3.72~\AA~our revised calibration implies an iron-abundance, stellar
metallicity of 0.267 solar.

\subsection{Fitting R04 model galaxy spectra}\label{cfmodels}

We independently determined the stellar metallicity by fitting the
five R04 spectra (for metallicities of 0.05,~0.2,~0.4,~1.0,~and~2.0
solar), broadened to match our rest-frame spectral resolution, to our
coadded observed spectrum for the wavelength range 1935-2020~\AA. In
our so-called ``renormalised method'' we decided the best-fit R04
model spectrum by minimising the median of the quadrature of
difference in flux between each broadened R04 model spectrum and our
coadded spectrum. We minimised the median flux difference by
multiplying the model spectrum by numbers between 0.9 and 1.1. We
searched for the minimum of the quadrature of the median and not of
the average because (i) the median is less sensitive than the average
to deviant pixels, and (ii) we could not reliably measure error as a
function of wavelength in our coadded spectrum (although to first
order this should be fairly constant for our wavelength range of
interest here). In our direct method we fitted each broadened R04
model spectrum without any multiplication by a constant. In Figure
\protect\ref{r04model} we plot the quadrature of the median difference
in flux for the two different fitting methods at each R04 model
metallicity. Using both methods we find that the $Z$ = 0.2 solar R04
broadened model spectrum provides the best description of our coadded
spectrum. This is in good agreement with the iron-abundance, stellar
metallicity measured using the 1978~\AA\ index.

\subsection{Robustness of the measurement and error}

We assess the integrity of our stellar metallicity measurement and its
error in different ways.

On close examination of Figure~\ref{r04model} it is seen that the
faint absorption lines of our coadded spectrum are well reproduced by
the R04 model spectra. This provides reassurance that the model
spectra, broadened to match our spectra resolution, can be used to
measure the galaxy stellar metallicity of our data.

The agreement between the two independent, best-fitting methods,
completed in Section \ref{cfmodels}, implies that we are correctly and
consistently assessing the continuum level of the coadded
spectrum. This is critical because incorrect determination of the
local (pseudo-)continua levels can be the main source of random and
systematic error in the measurement of absorption-line equivalent
widths.

We further remark that when applying the renormalised method in
Section \ref{cfmodels} the constant required to minimise the flux
difference between coadded spectrum and model spectra (in cases of the
0.05, 0.2 and 0.4 solar metallicity models), modified the continuum
level of the model spectrum by typically 0.6\%. The coadded spectrum
continuum level is being reproduced to at least this level. A
conservative 1978~\AA\ index error measurement would be
$\sim$0.5~\AA~(the error corresponding to a shift of 0.6\% in the
level of the normalised coadded spectrum continuum).

\begin{figure} 
\centering
\includegraphics[width=9.0cm]{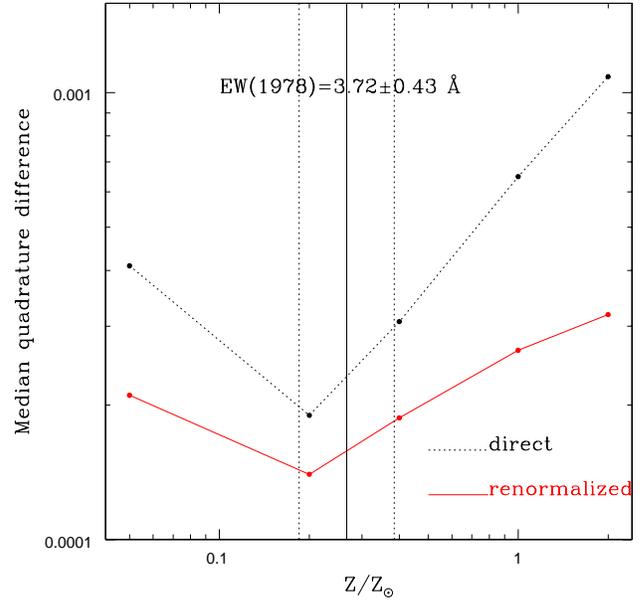}
\caption{\label{chisq_model} For the wavelength range 1935-2020~\AA~we
plot the quadrature of the median of flux difference between our
coadded spectrum and each R04 model spectrum broadened to match our
rest-frame spectral resolution. These differences were calculated as
described in Section \protect\ref{cfmodels}. The median is shown as a
function of R04 model spectrum metallicity in units of solar
metallicity for both the direct and renormalised methods. For both
methods the difference between our coadded spectrum and model spectra
is a minimum for the R04 model spectrum of 0.2 solar metallicity. The
dotted vertical lines indicate our stellar metallicity measurement and
error derived in Section \protect\ref{index}.}
\end{figure}

\begin{figure*} 
\centering
\includegraphics[width=20.0cm]{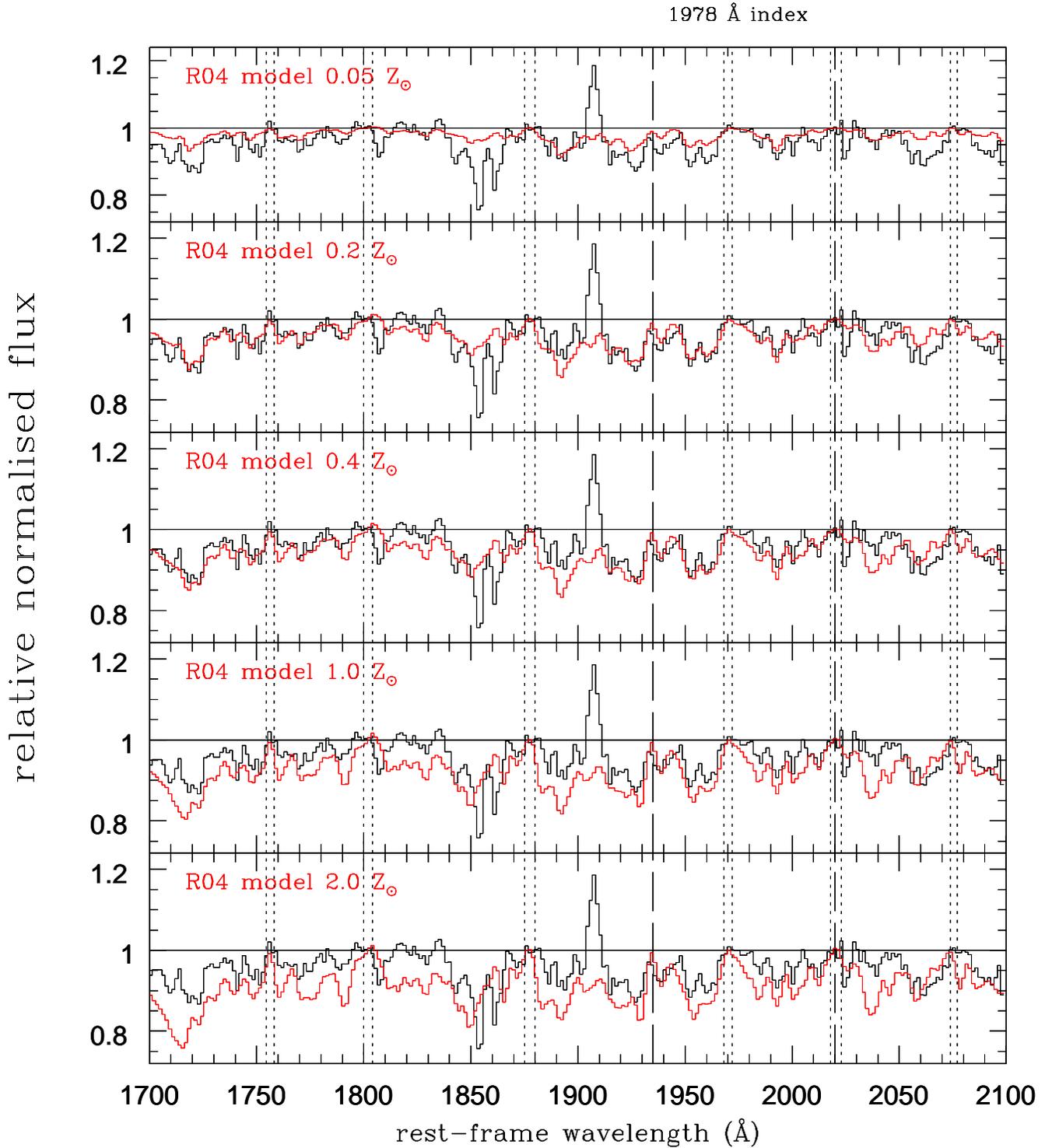}
\caption{\label{r04model} Plots of optimised best-fitting of our
coadded spectrum with the five R04 model galaxy spectra broadened to
match our spectral resolution. In each plot our coadded spectrum is
shown by a black solid line. Broadened R04 model spectra are
overplotted with red lines for metallicities = 0.05, 0.2, 0.4, 1.0,
and 2.0. Black long-dashed vertical lines indicate the wavelength
interval corresponding to the R04 1978~\protect\AA~index, and black
solid horizontal lines delineate a normalised spectrum flux level of
1.0 .}
\end{figure*}

In Section \ref{cfmodels} and Figure \ref{chisq_model} the quadrature
of the median difference between the coadded spectrum and its best-fit
model spectrum, corresponds to $\sim$1.2\% of the coadded spectrum
continuum level. This implies that the assessment of the continuum in
the coadded spectrum (normalised using 6 R04 pseudo-continua
intervals) is accurate to a level of $\sim$0.5\%. Conversely the R04
model spectra are proven to reproduce an observed star-forming galaxy
spectrum at the level of $\simeq$1\%.

Using the above results we conservatively assume that the noise of
each bin of our coadded spectrum is $\sim$1\%~(implying a
signal-to-noise ratio S/N of 100 per 1.5~\AA\ rest-frame resolution
element).

We determined the 1978~\AA\ index error by simulating galaxy spectra
affected by noise. As previously noted a major source of error in the
1978~\AA\ index EW measurement is inaccurate assessment of the
continuum level using local pseudocontinua either side of the main
index passband. 1000 noise-affected galaxy spectra were created by
adding random noise to the mean flux in each of the 6 R04
pseudocontinua wavelength intervals. The 1978~\AA~index was measured
in each spectrum as described in Section \ref{index}. The standard
deviation of the 1000 1978~\AA~index measurements was 0.43~\AA. This
was taken to be our index error and is very close to our independent
assessments above. Using error propagation in Equation \ref{equ} this
error corresponds to an uncertainty in metallicity of 0.159~dex: this
is the metallicity errorbar plotted for our data in Figure
\ref{dave_models}.

Our measurement of stellar metallicity, derived using the
1978~\AA~index, and its corresponding error is 0.267
$_{-0.082}^{+0.118}$ solar (log$(Z/Z_\odot) = -0.574\pm0.159$).


\section{Discussion}\label{discuss}

\begin{figure*} 
\centering
\includegraphics[width=15.0cm]{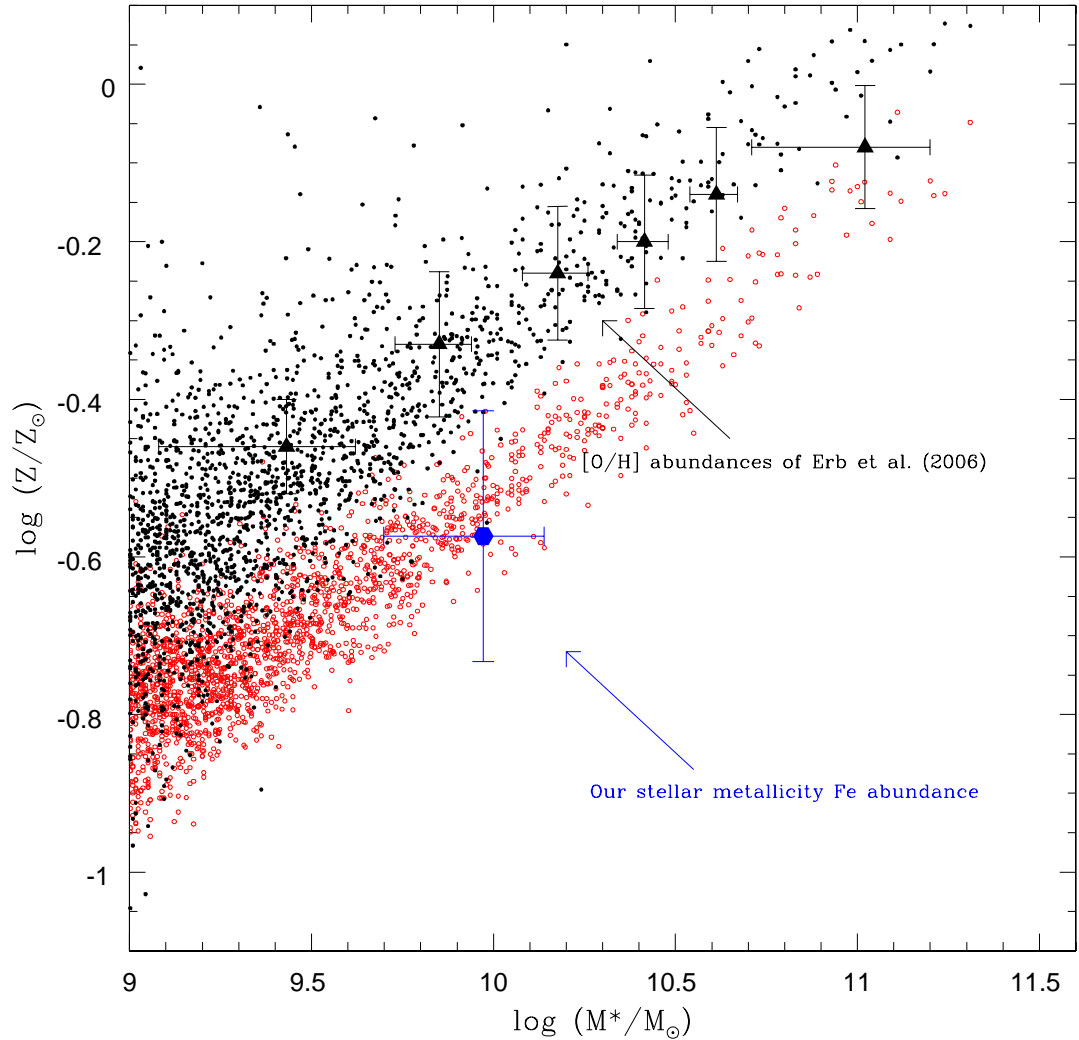}
\caption{\label{dave_models} The blue hexagon filled symbol indicates
our measurements of galaxy stellar mass and metallicity. Red circle
unfilled and black circle filled symbols provide cosmological model
predictions of galaxy stellar and gas metallicity respectively at
redshift z$\sim$2~from \protect\citet{fin07}. Black triangle filled
symbols are the gas-phase metallicity data of \protect\citet{erb06} at
redshift z$\sim$2.}
\end{figure*}

In Figure \ref{dave_models} we compare our measurement of galaxy
iron-abundance, stellar metallicity, with the oxygen-abundance,
gas-phase metallicity data of UV-selected star-forming galaxies from
Erb et al.~(2006, hereinafter E06), and the theoretical predictions of
both galaxy gas-phase and stellar metallicity for different galaxy
stellar masses from the cosmological simulations of \citet{fin07}
(hereinafter FD07).The estimated error of the N2 method itself as
indicated in Figure 3 of \cite{erb06} is approximately 0.05 dex. The
solar oxygen abundance measurement of \cite{asp04} has been subtracted
in log-space from the oxygen-abundance gas-phase metallicities of E06;
the stellar masses presented in E06 were calculated by assuming a
Chabrier IMF.

Before venturing into interpreting Figure \ref{dave_models} we need to
clarify that ``metallicity'' has a different meaning as measured by
us, observed by E06 or calculated by FD07. In our case, the
1978~\AA~index measures a photospheric, Fe~III absorption line of
massive OB stars, i.e., of the most recent stars to have formed out of
the ISM. The E06 measurement instead refers to the gas-phase oxygen
abundance for UV-selected star-forming galaxies using the N2
method. Finally, the FD07 ``metallicity'' refers to the mass-averaged
abundance of all the stars formed during the whole previous history of
each galaxy, and corresponds to all heavy elements having assumed a
global yield ($y=0.02$). Thus the three metallicities refers to three
different galactic components, and to three different elements or
combination of elements. Moreover, the three metallicity
determinations are derived from completely different procedures, each
affected by its own different systematic errors. With this caveat in
mind, we avoid overinterpreting the differences seen in Figure
\ref{dave_models}.

The $\sim$0.2 dex offset between the E06 oxygen abundance of the ISM
and the average metal abundance of the stars of the FD07 simulated
galaxies is qualitatively in agreement with the expectation. Indeed,
the ISM metallicity results from the whole past metal enrichment,
while the average stellar metallicity includes all previous, more
metal poor stellar generations.

More intriguing is the difference between our estimate and that of
either the ISM ($\sim$0.25 dex) or the stars in the simulated galaxies
($\sim$0.05 dex). Since we measure the metallicity in massive OB
stars, one would expect it to be nearly identical to that of the ISM,
given the short time elapsed since these stars formed (less than
$\sim$10 Myr). However, assuming that this difference is real and not
the effect of systematic error in the different methods, we must
consider that the ISM metallicity is a measure of oxygen abundance,
and our stellar metallicity is a measure of iron. As is well known,
oxygen is produced only by massive stars exploding as Type II
supernovae within at most few 10 Myr after formation, while iron is
also produced by Type Ia supernovae that are characterized by an
extremely wide distribution of formation-to-explosion delay times,
from $\sim$10 Myr to more than 10 Gyr
(e.g.~\citealt{gre05,man05}).

This different time behaviour of the two supernova types is understood
to be responsible for the $\alpha$ element enhancement (including the
abundance of oxygen) seen in stellar systems that formed over a time
short enough to correspond to a fraction of the SNIa distribution of
delay times. Indeed, such $\alpha$ element enhancements are observed
in the metal poor stars of the Galactic halo
(e.g. \citealt{whe89,cay04}), in metal poor and rich stars of the
Galactic bulge \citep{zoc06,ful07}, as well as in elliptical galaxies
(e.g. \citealt{dav93,kun98,hal99,kun00,kun01,mor04,tho05}). Since the
star-forming galaxies we are studying are undergoing a starburst
phase, the likely progeny of many if not most of them are bulges and
early-type galaxies which in the local universe exhibit an $\alpha$
enhancement, comparable to the 0.25 dex offset seen in Figure
\ref{dave_models}. Thus, taking this figure at face value, it suggests
that we may be witnessing the $\alpha$ enhancement being established
among these galaxies.

The original low-redshift mass-metallicity relation of T04 lies
$\sim$0.5 dex above the relation of E06. E06 also remeasured the
oxygen-abundance, gas-phase metallicity using the N2 method for the
same 53,400 galaxies studied by T04. They found a similar offset
although narrower at higher stellar mass close to solar metallicity as
the $[$NII$]$ emission line used in the N2 method begins to saturate.

We note that the weighting scheme implemented, and the requirement of
sufficient signal in the rest-frame UV to be able to measure a galaxy
redshift, is biasing our measurement against the most reddened $z=2$
galaxies (and thus likely the most metal-rich). Although the long
GMASS integration times of 15-20 hours are limiting this effect, still
it might be expected that our result could be somewhat underestimating
the true average metallicity of star-forming galaxies at the probed
stellar masses (we note that the data of \cite{erb06} probably suffer
an even stronger bias because of their UV-selection).

\section{Conclusions}\label{conclude}

We have measured the iron-abundance, stellar metallicity of
star-forming galaxies at redshift z$\sim$2, for a spectrum created by
adding together 75 star-forming galaxy spectra from the GMASS
survey. The iron-abundance, stellar metallicity is determined by
measuring the EW of a rest-frame mid-UV, photospheric absorption-line
index, the 1978~\AA\ index, defined by \cite{rix04} using the
theoretical massive star models of \cite{pau01}, and the evolutionary
population synthesis code Starburst99. We measure an iron-abundance,
stellar metallicity of 0.267 $_{-0.082}^{+0.118}$ solar
(log$(Z/Z_\odot) = -0.574\pm0.159$). This is lower by $\sim$0.25 dex
than the oxygen-abundance, gas-phase metallicity measured by
\citet{erb06} for similar galaxy stellar mass and redshift. At least
part of this difference may be the result of different systematic
errors between the two estimates although both ourselves and
\cite{erb06} find that systematic errors are very unlikely to
reproduce our observed data results. We postulate that our
measurements are reminiscent of the $\alpha$-element enhancement seen
in the likely progenitors of these starburst galaxies, i.e. galactic
bulges and early-type galaxies.

\begin{acknowledgements}

This paper is dedicated to the memory of Professor Mario Perinotto. We
thank the anonymous referee for constructive comments that helped
improve the clarity of the manuscript. We are grateful to Max Pettini
and Samantha Rix for providing the R04 models in digital form. We
thank Kristian Finlator and Romeel Dav\'e for providing their model
predictions for the stellar and gas metallicity of $z=2$ galaxies. CH
gratefully acknowledges the support and resources of IT Integration
Engineering and the hospitality and resources of the University of
Glasgow. ED is grateful to the Arcetri Observatory for hospitality
during the development of this work.

\end{acknowledgements}

\bibliographystyle{aa} \bibliography{8673chal}

\end{document}